\documentclass[preprint]{aastex}
\begin{document}
\title{Bulk flows and CMB dipole anisotropy \\
in cosmological void models}
\author{Kenji Tomita}
\affil{Yukawa Institute for Theoretical Physics, Kyoto University,
        Kyoto 606-8502, Japan}
\email{tomita@yukawa.kyoto-u.ac.jp}
\begin{abstract}
The observational behavior of spherically symmetric inhomogeneous
cosmological models is studied, which consist of inner and
outer homogeneous regions connected by a shell or an intermediate
self-similar region. It is assumed that the present matter density
parameter in the inner region is smaller than that in the outer
region, and the present Hubble parameter in the inner region is 
larger than that in the outer region. Then galaxies in the inner 
void-like region can be seen to have a bulk motion 
relative to matter in the outer region, when we observe them at a
point O deviated from the center C of the inner region. Their velocity
$v_p$ in the CD direction is equal to the difference of two Hubble 
parameters multiplied by the distance between C and O.
It is found also that the velocity $v_d$ corresponding to CMB dipole 
anisotropy observed at O is by a factor $\approx 10$ small compared with
$v_p$. This behavior of
$v_d$ and $v_p$ may explain the puzzling situation of the cosmic flow
of cluster galaxies, when the radius of the inner region and the
distance CD are about 200 $h^{-1}$ Mpc and 40 $h^{-1}$ Mpc,
respectively ($H_0 = 100 h^{-1}$ km sec$^{-1}$ Mpc$^{-1}$), and when the
gaps of density and Hubble parameters are $\approx 0.5$ and $18 \%$,
respectively.
\end{abstract}


\section{Introduction}
\label{sec:level1}

The dipole moment in the cosmic background radiation (CMB) is thought
to come mainly from the Doppler shift due to the motion of the Local
Group (LG), relative to the cosmic homogeneous expansion. As the main
gravitational source which brings the velocity vector of LG, the
existence of the Great Attractor (GA) was found by \citet{lyn88} and 
\citet{dress87}. It has the
position at the redshift of 4300 km sec$^{-1}$. On the other hand, the 
motion of LG in the inertial frame consisting of many clusters on
larger-scales was studied observationally by several groups: A
bulk flow of $\sim 700$ km sec$^{-1}$ was found by \citet{lp94}
and  \citet{coll95} as the motion of the Abell cluster inertial frame
relative to LG in the region with redshift $< 15000$ km sec$^{-1}$,
but in the other approach the different result was derived by \citet{giov98}, 
\citet{dale99} and \citet{riess97} in the regions with similar redshifts.
The Lauer and Postman's work is based on the assumption that the
brightest cluster galaxies as standard candles and the Hoessel
relation can be used, but at present these assumptions have been
regarded as questionable or unreliable.

Independently of these works, the motion of cluster frames relative to 
CMB was measured by \citet{hud99} and \citet{will99} due to the global 
Hubble formula
using the Tully-Fisher distances of clusters and their redshifts with
respect to CMB, and the flow velocity vector was derived in the region 
with about 150$h^{-1}$ Mpc  \ ($H_0 = 100 h^{-1}$ km sec$^{-1}$ 
Mpc$^{-1}$).  The remarkable and puzzling properties of these flows 
are that the flow velocity
reaches a large value $\sim 700$ km/sec on large scale, while the
dipole velocity (not due to GA) determined in the form of CMB dipole 
anisotropy seems to be much small, compared with the above flow velocity. 

If the observed large-scale matter motion is caused by the attraction 
from an over-density region containing superclusters, the
corresponding  
velocity must be so large as the large-scale flow velocity and it must 
be reflected in the form of CMB dipole anisotropy. 
If this motion is caused in the spherical void-like region, however,
the situation is different, because CMB dipole anisotropy near the
center can be relatively small in spite of the large-scale flow, 
as was shown in our previous paper \citep{tom96}. 
In this previous paper an inhomogeneous model 
on super-horizon scale was considered to explain the number evolution
of QSOs \citep{tom95}, but the relative smallness of the dipole 
anisotropy can be found independently of the scale of inhomogeneities. 
The local void region was studied independently by \citet{zehav98} as
the local Hubble bubble, which has the scale $\sim 70 h^{-1}$ Mpc and
is bordered by the dense walls. They analyzed the statistical relation 
between the distances and the local Hubble constants derived from the
data of SNIa, and found the existence of a void region with a local
Hubble constant larger than the global Hubble constant. The relation
to the SNIa data on larger scales will be discussed from our
standpoint in a subsequent paper.

In the present paper we consider more realistic inhomogeneous models on
sub-horizon scale, corresponding to matter flows $\sim 150 h^{-1}$
Mpc, which may be associated with large-scale structures or excess 
powers observed by \citet{broad90}, \citet{land96}, and \citet{ein97}.  
In \S \ref{sec:level2}, we treat a spherically symmetric
inhomogeneous model which
consists of inner and outer homogeneous regions connected by a shell
being a singular layer, and study the behavior of large-scale motions
caused in the inner region, where the present inner density parameter 
is smaller than the present outer density parameter and the present 
Hubble parameter in the inner region is larger than that in the outer 
region (a bulk motion in the void-like region was discussed also by
\citet{nak95}). In this section we treat the single-shell case.
The double-shell case and a model with an intermediate self-similar
region are treated in Appendixes A and B, respectively.
In \S \ref{sec:level3}, we consider light rays which are emitted at
the last scattering surface and reach an observer situated at a point
O deviated from the center C, and CMB dipole and quadrupole
anisotropies are analyzed. The peculiar velocity of the above
large-scale motions and the velocity corresponding to the CMB dipole
anisotropy are compared. In \S \ref{sec:level4}, a naive explanation 
about why CMB dipole anisotropy is small around the center C is shown. 
In \S \ref{sec:level5} the consistency of the present models with
several recent observations of bulk flows is discussed, and in \S 
\ref{sec:level6} concluding remarks are presented.

\section{Inhomogeneous models and the bulk motion}
\label{sec:level2}

In this section we consider  spherically symmetric inhomogeneous
cosmological models which have two homogeneous regions connected with a
spherical shell, as shown in Fig. \ref{fig1}.

The line-elements in the inner region V$^{\rm I}$ and the outer region 
V$^{\rm II}$ 
are described as
\begin{equation}
  \label{eq:m1}
 ds^2 = g^j_{\mu\nu} (dx^j)^\mu (dx^j)^\nu = - c^2 (dt^j)^2 + [a^j
 (t^j)]^2 \Big\{ d (\chi^j)^2 + [f^j (\chi^j)]^2 d\Omega^2 \Big\},
\end{equation}
where $j \ (=$ I or II) represents the regions, $f^j (\chi^j) = \sin
\chi^j, \chi^j$ and $\sinh \chi^j$ for $k^j = 1, 0, -1$, respectively,
and $d\Omega^2 = d\theta^2 + \sin^2 \theta \varphi^2$.
The shell is a time-like hypersurface $\Sigma$ given as $\chi^{\rm I} =
\chi^{\rm I}_1$ and $\chi^{\rm II} = \chi^{\rm II}_1$.

\subsection{Cosmological models}
The Einstein equations are divided into the equations in the two regions
and the jump conditions at the shell. 
The general formulation of the
jump condition at the singular surface was derived by \citet{isr66} and the
concrete expressions of conditions were derived by \citet{maed86} 
 and \citet{sak93}. Here the expressions by Sakai et al. 
are shown using the
circumferential radius of the shell $R$, the velocity of the shell
$v^j$, the Lorentz factor $\gamma^j$ and the Hubble expansion
parameter $H^j$ in V$^j$ \ ($j =$ I and II) defined by

\begin{equation}
  \label{eq:m2}
R \equiv a^{\rm I} f^{\rm I} = a^{\rm II} f^{\rm II}, \ v^j \equiv
a_j{d\chi^j \over dt^j}, \ \gamma^j \equiv 1/\sqrt{1 - \Big({v^j 
\over c}\Big)^2} \ {\rm and} \ H^j \equiv {da^j/dt^j \over a^j}. 
\end{equation}
The Einstein equations for the pressureless matter in the two regions
are 
\begin{equation}
  \label{eq:m3}
(H^j)^2 + k^j c^2/(a^j)^2 = {8\pi G\over 3} \rho^j + {1 \over 3}
\Lambda c^2,
\end{equation}
where $\rho^j$ is the mass density of matter  $(\propto 1/(a^j)^3)$.

The equations for the surface density $\sigma$ and the velocity
$v^{\rm II}$ of the shell are expressed as
\begin{equation}
  \label{eq:m4}
\gamma^{\rm II} d(4\pi R^2\sigma)/dt^{\rm II} = \Big[4\pi R^2 
\gamma^2 v\rho \Big]^{\rm II,I}, 
\end{equation}
\begin{equation}
  \label{eq:m5}
d(\gamma^{\rm II} v^{\rm II})/dt^{\rm II} = -\gamma^{\rm II} v^{\rm
II} H^{\rm II} + 2\pi G\sigma - [\gamma^2 v^2 \rho]^{\rm II}/\sigma,
\end{equation}
where $[\Phi]^{\rm II,I} \equiv \Phi^{\rm II} - \Phi^{\rm I}$.
The conditions of continuity of the metric $(d\tau^2 = -ds^2)$ and the 
common velocity $dR/d\tau$ reduce to
\begin{equation}
  \label{eq:m6}
dt^{\rm I}/dt^{\rm II} = \gamma^{\rm I}/\gamma^{\rm II}
\end{equation}
and
\begin{equation}
  \label{eq:m7}
\gamma^{\rm I}(f'^{\rm I} v^{\rm I} + H^{\rm I}R) = \gamma^{\rm II}
(f'^{\rm II} v^{\rm II} + H^{\rm II} R),
\end{equation}
where $f'^j = df^j (\chi^j)/d\chi^j$.

Another important component of jump conditions playing a role of a
constraint equation is
\begin{equation}
  \label{eq:m8}
[\gamma(f' +vHR)]^{\rm II,I} = - 4\pi G\sigma R.
\end{equation}
Solving Eqs.(\ref{eq:m4}) and (\ref{eq:m5}) we can obtain the time
evolution of $\sigma$ and $v_{\rm II}$ in the shell, and $v^{\rm I}$
is derived using Eq.(\ref{eq:m7}). These values of $\sigma, v^{\rm
I}$ and $v^{\rm II}$ satisfy the condition (\ref{eq:m8}). The
initial condition is given as a form of $(H^{\rm I})_i = (H^{\rm II})_i$
at an initial epoch $(t^j)_i$ such as the recombination epoch.

The background models in V$^{\rm I}$ and V$^{\rm II}$ are rewritten
using 
\begin{equation}
  \label{eq:m9}
y^j \equiv a^j/(a_0)^j, \  \tau^j \equiv H_0^j t^j, \
\lambda_0^j \equiv {1 \over 3} \Lambda c^2/(H_0^j)^2, \ \Omega_0^j
\equiv {8\pi G \over 3(H_0^j)^2} (\rho_0)^j,
\end{equation}
where $0$ denotes the present epoch. Eq. (\ref{eq:m3}) is given as
\begin{equation}
  \label{eq:m10}
dy^j/d\tau^j = (y^j)^{-1/2} P_j (y^j),  
\end{equation}
where
\begin{equation}
  \label{eq:m11}
\ P_j (y^j)\equiv [\Omega_0^j +
\lambda_0^j (y^j)^3 + (1 - \Omega_0^j - \lambda_0^j) y^j]^{1/2}
\end{equation}
and $(a_0)^j$ is given by
\begin{equation}
  \label{eq:m12}
(a_0 H_0)^j = 1/\sqrt{1 - \Omega_0^j - \lambda_0^j}.
\end{equation}
The conformal times $\eta^j$ are defined by
\begin{equation}
  \label{eq:m13}
\eta^j \equiv \sqrt{1 - \Omega_0^j - \lambda_0^j} \int_0^{y^j} dy/[y^{1/2} 
P^j (y)].
\end{equation}
The solution of Eq. (\ref{eq:m10}) in the case of $k^{\rm I} = k^{\rm
II} = -1$ and $\Lambda = 0$ are expressed as
\begin{equation}
  \label{eq:m14}
y^j = {\Omega_0^j \over 2(1 - \Omega_0^j)} (\cosh \eta^j -1), 
\end{equation}
\begin{equation}
  \label{eq:m15}
\tau^j = {\Omega_0^j \over 2(1 - \Omega_0^j)^{3/2}} (\sinh \eta^j -
\eta^j).
\end{equation}
In the case of nonzero $\Lambda$, \  Eq. (\ref{eq:m10}) is solved
numerically.

Eqs. (\ref{eq:m4}) and (\ref{eq:m5}) were solved by \citet{sak93} and
it was shown in the case of $\Omega_i \simeq 1$ that the present value 
of $v^{\rm II}$ at $a/a_i > 100$ is $< 100$ km/s, as long as the
shell starts with the vanishing initial velocity $(v^j)_i = 0$.

Here the initial condition $(H^{\rm I})_i = (H^{\rm II})_i$ at 
initial epoch $y^j = (y^j)_i$ is expressed as
\begin{eqnarray}
  \label{eq:m16}
H_0^{\rm I} &&(y_i^{\rm I})^{-3/2} [\Omega_0^{\rm I} + \lambda_0^{\rm I} 
(y_i^{\rm I})^3 + (1 - \Omega_0^{\rm I} - \lambda_0^{\rm I}) 
y_i^{\rm I}]^{1/2} \cr
&&= H_0^{\rm II} (y_i^{\rm II})^{-3/2} 
[\Omega_0^{\rm II} + \lambda_0^{\rm II} (y_i^{\rm II})^3 + (1 - 
\Omega_0^{\rm II} - \lambda_0^{\rm II}) y_i^{\rm II}]^{1/2},
\end{eqnarray}
where we have
\begin{equation}
  \label{eq:m17}
{\Omega_0^{\rm II} \over \Omega_0^{\rm I}} = {\rho_0^{\rm II} \over 
\rho_0^{\rm I}} \Big({H_0^{\rm I} \over 
H_0^{\rm II}}\Big)^2, \quad {\lambda_0^{\rm II} \over \lambda_0^{\rm I}} 
= \Big({H_0^{\rm I} \over H_0^{\rm II}}\Big)^2
\end{equation}
from Eq. (\ref{eq:m9}). If we eliminate $\Omega_0^{\rm II}, 
\lambda_0^{\rm II}$ from Eq. (\ref{eq:m16}) using
Eq. (\ref{eq:m17}), we obtain 
\begin{equation}
  \label{eq:m18}
\Big[{\rho_0^{\rm II} \over \rho_0^{\rm I}} (1 - y_i^{\rm II}) - 
\Big({y_i^{\rm II} \over y_i^{\rm I}}\Big)^3 (1 - y_i^{\rm I})\Big] 
\Omega_0^{\rm I} - \Big[1 - \Big({y_i^{\rm II} \over 
y_i}\Big)^2\Big] y_i^{\rm II} \lambda_0^{\rm I} + \Big[\Big(
{H_0^{\rm II} \over H_0^{\rm I}}\Big)^2 - \Big({y_i^{\rm II} 
\over y_i^{\rm I}}\Big)^3 \Big] y_i^{\rm II} = 0.
\end{equation}
Since $y_i^{\rm I} \sim y_i^{\rm II} \sim 10^{-3} \ (<< 1)$, we assume 
\begin{equation}
  \label{eq:m19}
{\rho_0^{\rm II} \over \rho_0^{\rm I}} = \Big({y_i^{\rm II} 
\over y_i^{\rm I}}\Big)^3 (1 + \epsilon y_i^{\rm I}),
\end{equation}
where $\epsilon \approx 1$. Then we get from Eq. (\ref{eq:m18})
\begin{equation}
  \label{eq:m20}
\Big({H_0^{\rm II} \over H_0^{\rm I}}\Big)^2 = \Big({y_i^{\rm II} 
\over y_i^{\rm I}}\Big)^3 +  \Big[1 - \Big({y_i^{\rm II} 
\over y_i^{\rm I}}\Big)^2 \Big]\lambda_0^{\rm I} + \Big({y_i^{\rm II} 
\over y_i^{\rm I}}\Big)^3  \Big[1 - (1 + \epsilon) {y_i^{\rm I} 
\over y_i^{\rm II}}\Big] \Omega_0^{\rm I}.
\end{equation}
If we give $\Omega_0^{\rm I}, \lambda_0^{\rm I}, \epsilon$ and 
${y_i^{\rm II} / y_i^{\rm I}}$,
therefore, we can obtain $H_0^{\rm II}/H_0^{\rm I}$ for
Eq. (\ref{eq:m20}) and derive ${\rho_0^{\rm II} / \rho_0^{\rm I}}$ 
from Eq. (\ref{eq:m20}) and ${\Omega_0^{\rm II} / \Omega_0^{\rm
I}}$ and ${\lambda_0^{\rm II} / \lambda_0^{\rm I}}$ from Eq. 
(\ref{eq:m17}). From Eqs. (\ref{eq:m17}), (\ref{eq:m19}) and 
(\ref{eq:m20}), we find that $H_0^{\rm II}/H_0^{\rm I} < 1$ and
${\Omega_0^{\rm II} / \Omega_0^{\rm I}} > 1$, if
\begin{equation}
  \label{eq:m20a}
1 > {y_i^{\rm II} \over y_i^{\rm I}} >  1 +
 \Big[1 - \Big({y_i^{\rm II} \over y_i^{\rm I}}\Big)^2 \Big] 
\Big({y_i^{\rm I} \over y_i^{\rm II}}\Big)^3
\lambda_0^{\rm I} +   \Big[1 - (1 + \epsilon) {y_i^{\rm I} 
\over y_i^{\rm II}}\Big] \Omega_0^{\rm I}.
\end{equation}

In the case $\Lambda = 0$, we have an example for $\Omega_0^{\rm I} = 0.2$
\begin{equation}
  \label{eq:m21}
\Omega_0^{\rm II} = 0.56, \ H_0^{\rm II}/H_0^{\rm I} = 0.80, \
{\rho_0^{\rm II} / \rho_0^{\rm I}} = 1.8, \ y_i^{\rm II} / y_i^{\rm
I} = 1.2, \  \epsilon = 4.1.
\end{equation}
In the case $\Lambda \neq 0$, we have an example
\begin{equation}
  \label{eq:m22}
\lambda_0^{\rm I}=0.672, \ \lambda_0^{\rm II}= 0.43, \ \Omega_0^{\rm
I} =0.3, \ \Omega_0^{\rm II} = 0.563, \ H_0^{\rm II}/H_0^{\rm I} =
0.80, \ \epsilon = 0.64, 
\end{equation}
so that 
\begin{equation}
  \label{eq:m23}
\Omega_0^{\rm I}+\lambda_0^{\rm I}=0.872, \quad \Omega_0^{\rm II}
+\lambda_0^{\rm II}=0.993.
\end{equation}

\subsection{Bulk motion}
Now let us consider the velocity field around an observer O in V$^{\rm
I}$ at the point with $l_0 \equiv (a\chi)_0 << (a\chi)_1$. Since $l_0$ is 
much smaller than the curvature radius, we can approximately neglect
the spatial curvature around him. Then he has the relative velocity
\begin{equation}
  \label{eq:m24}
\Delta v_0 = (H_0^{\rm I} - H_0^{\rm II}) l_0 
\end{equation}
to matter in the outer region in the direction of the $X$ axis.
 If $H_0^{\rm I} = 100 h$ km/sec/Mpc, \ 
$H_0^{\rm II} = 0.82 H_0^{\rm I}$ and $l_0 = 40 h^{-1}$ Mpc, we have 
$\Delta v_0 = 720$ km/sec. 

Here consider a galaxy G with the radius coordinate $\chi$ and angle 
$\varphi$. Then the relative velocity of G to matter in the outer region is 
\begin{equation}
  \label{eq:m25}
\Delta v_G = (H_0^{\rm I} - H_0^{\rm II}) (a_0 \chi)
\end{equation}
in the radial direction from the center C of the inner region
(cf. Fig. \ref{fig3}).

This velocity can be divided into the $X$ component $(\Delta v_G)_X$ 
and the line-of-sight component $(\Delta v_G)_{\rm LS}$ with respect to 
the observer O as follows, noticing that the angle $\angle$GOX \ ($=
\phi$) satisfies the relation 
\begin{equation}
  \label{eq:m26}
\sin \phi = \chi \sin \varphi/[\chi^2+\chi_0^2 -2\chi\chi_0 \cos
\varphi]^{1/2}, \quad \cot \phi = (\cos \varphi -\chi_0/\chi)/\sin \varphi.
\end{equation}
Their values are
\begin{equation}
  \label{eq:m27}
(\Delta v_G)_X = \Delta v_G \sin (\phi - \varphi)/\sin \phi = \Delta
v_0,
\end{equation}
\begin{equation}
  \label{eq:m28}
(\Delta v_G)_{LS} = \Delta v_G \sin \varphi/\sin \phi = \Delta v_G [1
- 2(\chi_0/\chi)\cos \varphi + (\chi_0/\chi)^2]^{1/2} \simeq \Delta v_G.
\end{equation}
That is, the $X$ component $(\Delta v_G)_X$ is constant and equal
to $\Delta v_0$, and another component, which is in the line-of-sight
direction from the observer, is nearly equal to $\Delta v_G$. Because
the first component $(\Delta v_G)_X$ is independent of the position, it 
can be interpreted as the peculiar velocity $v_p$ of galaxies which
represents the bulk motion:
\begin{equation}
  \label{eq:m29}
v_p = (\Delta v_G)_X = \Delta v_0.
\end{equation}

\section{Redshift formula and the CMB anisotropy}
\label{sec:level3}
 The wavevector $k^\mu$ in the inner and outer regions in the plane of 
 $\theta = \pi/2$ is obtained by solving null-geodesic equation 
and expressed as 
\begin{equation}
  \label{eq:r1}
(k^0)^j = dt^j/d\lambda = {a_0^j \over a^j} w_1^j,
\end{equation}
\begin{equation}
  \label{eq:r2}
(k^\chi)^j = d\chi^j/d\lambda = \pm {a_0^j w_1^j \over (a^j)^2} \Big[1 - 
\Big({d^j/w_1^j \over a_0^j \sinh \chi^j}\Big)^2\Big]^{1/2},
\end{equation}
\begin{equation}
  \label{eq:r3}
(k^\varphi)^j = d\varphi/ d\lambda = d^j/\Big(a^j \sinh \chi^j \Big)^2,
\end{equation}
where $j =$ I and II, and $\lambda$ is an affine parameter. 

\subsection{Light paths}

In the inner region V$^{\rm I}$, it is assumed that at the present epoch 
($t^{\rm I} = t_0^{\rm I}$) all rays reach an observer at the point O
with $\chi = \chi_0, \ \theta = \pi/2$ and $\varphi = 0$ in the $X$ axis,
and the angle between the rays and the $X$ axis is $\phi$. Then we have 

%
\begin{equation}
  \label{eq:r3a}
\phi = \phi_1  \quad {\rm and} \quad \pi - \phi_1, 
\end{equation}
where
\begin{equation}
  \label{eq:r4}
\phi_1 \equiv \sin^{-1} \Big[{d^{\rm I} /w_1^{\rm I} \over a_0^{\rm I} 
 \sinh \chi_0^{\rm I}} \Big] \ ( < \pi/2). 
\end{equation}
For $\phi = \phi_1$ the rays are expressed as
\begin{equation}
  \label{eq:r5}
G(\chi^{\rm I}) \equiv \cosh^{-1} \Big({\cosh \chi^{\rm I} \over
h_0^{\rm I}}\Big) - \cosh^{-1} \Big({\cosh \chi_0^{\rm I} \over
h_0^{\rm I}}\Big) = \eta_0^{\rm I} - \eta^{\rm I}, 
\end{equation}
where
\begin{equation}
  \label{eq:r6}
h_0^j \equiv \Big[1 + \Big({d^j/w_1^j \over (a_0)^j} \Big)^2
\Big]^{1/2},
\end{equation}
$\eta^{\rm I}$ is defined by Eqs. (\ref{eq:m13}) and (\ref{eq:m14}),
and $\eta_0^{\rm I}$ is equal to $\eta^{\rm I}$ at present epoch 
($y^{\rm I} = 1$). 

For $\phi = \pi - \phi_1$ we have
\begin{eqnarray}
  \label{eq:r7}
G(\chi^{\rm I}) &=& -\eta_0^{\rm I} +\eta^{\rm I}, \quad  {\rm for}\ 
\eta_0^{\rm I} \geq \eta^{\rm I}> \eta_m,  \cr
 &=&  -\eta^{\rm I} -\eta_0^{\rm I} + 2 \eta_m,  \quad  {\rm for}\ 
\eta^{\rm I} \leq \eta_m,
\end{eqnarray}
where 
\begin{equation}
  \label{eq:r8}
\eta_m \equiv \eta_0^{\rm I} - \cosh^{-1} \Big({\cosh \chi_0^{\rm I} \over
h_0^{\rm I}}\Big).
\end{equation}
\noindent
In the latter case, $\chi^{\rm I}$ has the minimum value (i.e., $k^{\chi} = 0$)
at $\eta^{\rm I} = \eta_m$.

At the boundary $\eta^{\rm I} = \eta_1^{\rm I}$ and $\chi^{\rm I} = 
\chi_1^{\rm I}$, therefore, we obtain 
\begin{equation}
  \label{eq:r9}
\eta_1^{\rm I} = \eta_0^{\rm I} \ \pm \ \cosh^{-1} \Big({\cosh \chi_0^{\rm
I} \over h_0^{\rm I}} \Big) - \cosh^{-1} \Big({\cosh \chi_1^{\rm
I} \over h_0^{\rm I}} \Big)
\end{equation}
for $\phi = \left(\matrix{\phi_1 \cr \pi - \phi_1}\right)$, respectively.
In the outer region V$^{\rm II}$, we have
\begin{equation}
  \label{eq:r10}
G(\chi^{\rm II}) \equiv \cosh^{-1} \Big({\cosh \chi^{\rm II} \over
h_0^{\rm II}}\Big) - \cosh^{-1} \Big({\cosh \chi_1^{\rm II} \over
h_0^{\rm II}}\Big) = \eta_1^{\rm II} - \eta^{\rm II},
\end{equation}
where $\eta^{\rm II}$ is given by Eqs. (\ref{eq:m13}) and
(\ref{eq:m14}), and $\eta_1^{\rm II}$ and $\chi_1^{\rm II}$ are the
values at the shell. At the recombination epoch we have
\begin{equation}
  \label{eq:r11}
\eta_{\rm rec}^{\rm II} = \eta_1^{\rm II} - \cosh^{-1} \Big({\cosh 
\chi_{\rm rec}^{\rm II} \over h_0^{\rm II}} \Big) + \cosh^{-1} \Big({\cosh 
\chi_1^{\rm II} \over h_0^{\rm II}} \Big).
\end{equation}
The junction of wavevectors at the boundary is expressed as
\begin{equation}
  \label{eq:r12}
(k^0)^{\rm I} = (k^0)^{\rm II},
\end{equation}
\begin{equation}
  \label{eq:r13}
(\sqrt{g_{\chi\chi}} \ k^\chi)^{\rm I} = (\sqrt{g_{\chi\chi}} \ 
k^\chi)^{\rm II},
\end{equation}
\begin{equation}
  \label{eq:r14}
(\sqrt{g_{\varphi\varphi}} \ k^\varphi)^{\rm I} = (\sqrt{g_{\varphi
\varphi}} \  k^\varphi)^{\rm II}.
\end{equation}
Eqs. (\ref{eq:r12}) and (\ref{eq:r14}) give
\begin{equation}
  \label{eq:r15}
\Big({a_0 \over a_1}\Big)^{\rm I} w_1^{\rm I} = \Big({a_0 
\over a_1}\Big)^{\rm II} w_1^{\rm II}, 
\end{equation}
\begin{equation}
  \label{eq:r16}
d^{\rm I} = d^{\rm II}, 
\end{equation}
respectively, where we used the relation $R = (af)^{\rm I} = (af)^{\rm
II}$. The conditions (\ref{eq:r12}) and (\ref{eq:r14}) are evidently
consistent with Eq. (\ref{eq:r13}), since $k^\mu$ is a null vector.

\subsection{Redshift formula}
Now let us derive the redshift formula for rays which are emitted at
the recombination epoch. Here this epoch is defined as the time of the 
radiation temperature $T_r = 10^3 (T_r)_0$ in the region V$^{\rm II}$,
where $(T_r)_0$ is the present temperature ($\simeq 2.7$ K). The total 
redshift factor $(1 + z_{\rm rec})$ is calculated as the product of two
redshift factors which are caused in the two regions V$^{\rm I}$ and
V$^{\rm II}$. 

First we assume that the shell is comoving and later the correction
due to the motion of the shell is examined. If we consider a virtual
observer at the center C \ $(\chi = 0)$, a light ray which is 
received by him at $\bar{\eta}^{\rm I} = \bar{\eta}_0^{\rm I}$ is expressed as
\begin{equation}
  \label{eq:r17}
\bar{\eta}_0^{\rm I} - \bar{\eta}^{\rm I} = \chi^{\rm I}, \ 
\bar{\eta}_0^{\rm I} - \bar{\eta}_1^{\rm I} = \chi_1^{\rm I}
\end{equation}
in V$^{\rm I}$, and 
\begin{equation}
  \label{eq:r18} 
\bar{\eta}_{\rm rec}^{\rm II} - \bar{\eta}^{\rm II} = \chi_{rec}^{\rm 
II} - \chi^{\rm II}, \ \bar{\eta}_{\rm rec}^{\rm II} -
\bar{\eta}_1^{\rm II} = \chi_{\rm rec}^{\rm II} - 
\chi_1^{\rm II}
\end{equation}
in V$^{\rm II}$, when the ray is emitted at the recombination epoch.

The redshift factors are 
\begin{equation}
  \label{eq:r19} 
1 + \bar{z}_1^{\rm I} = {a_0^{\rm I} \over a^{\rm I}(\bar{\eta}_1^{\rm I})} 
= {1 \over y^{\rm I}(\bar{\eta}_1^{\rm I})}
\end{equation}
in V$^{\rm I}$, and 
\begin{equation}
  \label{eq:r20}  
{1 + \bar{z}_{\rm rec}^{\rm II} \over 1 + \bar{z}_1^{\rm II}} = 
{a^{\rm II}(\bar{\eta}_1^{\rm II}) \over
a^{\rm II}(\bar{\eta}_{\rm rec}^{\rm II})} = {y^{\rm II}(\bar{\eta}_1^{\rm II})
\over y^{\rm II}(\bar{\eta}_{\rm rec}^{\rm II})}
\end{equation}
in V$^{\rm II}$.
The junction condition in Eq. (\ref{eq:m6}) gives $\bar{z}_1^{\rm I} = 
\bar{z}_1^{\rm II}$ for $v^{\rm I} = v^{\rm II} = 0$, so that 
\begin{equation}
  \label{eq:r21}
1 + \bar{z}_{\rm rec}^{\rm II} = {y^{\rm II}(\bar{\eta}_1^{\rm II}) \over
y^{\rm II}(\bar{\eta}_{\rm rec}^{\rm II})  y^{\rm I}(\bar{\eta}_1^{\rm I})}.
\end{equation}
Here we specify the value of $\bar{z}_1^{\rm I}$ as $\bar{z}_1^{\rm I} 
= 0.067 \sim 0.1$. Then $\bar{\eta}_1^{\rm I}$ and $\chi_1^{\rm I}$ are 
determined from Eqs. (\ref{eq:r17}) and (\ref{eq:r19}), and $\bar{\eta}_1^{\rm
II}$ and $\chi_1^{\rm II}$ are determined using the relations 
\begin{equation}
  \label{eq:r22}
a_0^{\rm I} y^{\rm I} (\bar{\eta}_1^{\rm I}) \sinh \chi_1^{\rm I} = a_0^{\rm
II} y^{\rm II} (\bar{\eta}_1^{\rm II}) \sinh \chi_1^{\rm II} 
\end{equation}
and 
\begin{equation}
  \label{eq:r23}
a_0^{\rm I} \int_0^{\bar{\eta}_1^{\rm I}} y^{\rm I} (\eta^{\rm I}) 
d\eta^{\rm I} = a_0^{\rm II} \int_0^{\bar{\eta}_1^{\rm II}} y^{\rm II} 
(\eta^{\rm II}) d\eta^{\rm II},
\end{equation}
which are obtained from Eqs. (\ref{eq:m2}) and (\ref{eq:m6}).
If $\bar{z}_{\rm rec}^{\rm II}$ is moreover specified (in the following we
take the value $\bar{z}_{\rm rec}^{\rm II} = 10^3 -1$), $y
(\bar{\eta}_{\rm rec})$ or $\bar{\eta}_{\rm rec}$ is determined 
using Eq. (\ref{eq:r21}). 

Next we consider the observer at O (with $\chi^{\rm I} = \chi_0^{\rm I}$ 
 and $\varphi = 0$). The above determined $\chi_1^j \ (j = {\rm I \ 
and \ II})$ and $\eta_{\rm rec}^{\rm II} \ (= \bar{\eta}_{\rm 
rec}^{\rm II})$ are used also for rays reaching this observer,
 to identify the position of the shell and the recombination epoch.
It should be noted that $\eta_{\rm rec}^{\rm II}$ 
depends only on the temperature at the recombination epoch and is
 independent of the existence of the inner region V$^{\rm I}$.

In V$^{I}$ we have the redshift
\begin{equation}
  \label{eq:r24}
1 + z_1^{\rm I} = {a_0^{\rm I} \over a^{\rm I}(\eta_1^{\rm I})} = 
{1 \over y^{\rm I} (\eta_1^{\rm I})}
\end{equation}
and in V$^{\rm II}$ we have 
\begin{equation}
  \label{eq:r25}
{1 + z_{\rm rec}^{\rm II} \over 1 + z_1^{\rm II}} = {a^{\rm II}
(\eta_1^{\rm II}) \over
a^{\rm II}(\eta_{\rm rec}^{\rm II})} = {y^{\rm II} 
(\eta_1^{\rm II}) \over y^{\rm II} (\eta_{\rm rec}^{\rm II})}.
\end{equation}
For a given $\chi_1^{\rm I}$, we obtain $\eta_1^{\rm I}$ by solving
Eq. (\ref{eq:r9}) and obtain $z_1^{\rm I}$ from Eq. (\ref{eq:r24}).
The junction condition in Eq. (\ref{eq:m6}) gives $z_1^{\rm I} = 
z_1^{\rm II}$, and $\eta_1^{\rm II}$ is related to $\eta_1^{\rm I}$
using the relations (\ref{eq:r22}) and (\ref{eq:r23}). Then the
product of Eqs. (\ref{eq:r24}) and (\ref{eq:r25}) reduces to
\begin{equation}
  \label{eq:r26}   
1 + z_{\rm rec}^{\rm II}  = {y^{\rm II} (\eta_1^{\rm II}) 
\over y^{\rm I} 
(\eta_1^{\rm I}) y^{\rm II} (\eta_{\rm rec}^{\rm II})},
\end{equation}
which is given as a function of angle $\phi$.

Now let us consider the case when the shell is not comoving, i.e., 
$v^{\rm I} \ne 0, \ v^{\rm II} \ne 0$. As was shown by Sakai et al.,
the velocities are $< 200$ km/sec and $(v^j/c)^2 < 10^{-7}$ for $j =$
I and II. Accordingly, we can assume $\gamma^j = 1$ for $j =$ I and
II, so that the condition $z_1^{\rm I} = z_1^{\rm II}$ and Eq. 
(\ref{eq:r23}) hold. In order to take the shell motion into account,
we assume the relation
\begin{equation}
  \label{eq:r27} 
\chi_1^{\rm I} = (\chi_1^{\rm I})_{\phi = 0} + v^{\rm I} [\eta_1^{\rm I} 
 - (\eta_1^{\rm I})_{\phi = 0}]
\end{equation}
in Eq. (\ref{eq:r9}) for arbitrary $\phi$, where $(\chi_1^{\rm 
I})_{\phi = 0}$ and 
$(\eta_1^{\rm I})_{\phi = 0}$ are their values for the ray incident in 
the direction of $\phi = 0$.
Subsequent calculations are same as the calculations for $v^{\rm I} = 0$.
 
The derivations of $z_{\rm rec}^{\rm II}$ in the case of double shells and 
in the inhomogeneous model with an intermediate self-similar region
are shown in Appendixes A and B.

\subsection{CMB anisotropy}
The values of $z_{\rm rec}^{\rm II}$ are numerically calculated for $0 <
\phi < \pi$, and their dipole and quadrupole moments are derived. 
When $z_{\rm rec} \ (= z_{\rm rec}^{\rm II} (\phi))$ is given, the temperature 
$T(\phi)$ of the cosmic background radiation is proportional to $1/(1
+ z_{\rm rec})$, and the dipole moment $D$ and quadrupole moment $Q$ are
defined as 
\begin{equation}
  \label{eq:r28}
D \equiv \Bigl\vert\int^\pi_0\int^{2\pi}_0 (1+z_{\rm rec})^{-1}Y_{10} \
\sin \phi d\phi d\varphi\Bigr\vert /<(1+z_{\rm rec})^{-1}>,
\end{equation}
\begin{equation}
  \label{eq:r29}
Q \equiv \Bigl\vert\int^\pi_0\int^{2\pi}_0 (1+z_{\rm rec})^{-1}Y_{20} \
\sin \phi d\phi d\varphi \Bigr\vert /<(1+z_{\rm rec})^{-1}>,
\end{equation}
where $< >$ means the average value taken over the whole sky 
, and
\begin{equation}
  \label{eq:r30}
Y_{10}(\phi) = \sqrt{3 \over 4\pi}\cos \phi \quad
Y_{20}(\phi) = \sqrt{5 \over 4\pi}\Bigl({3 \over 2}\cos^2 \phi - {1
\over 2}\Bigr).
\end{equation}
The Doppler velocity $v_d$ corresponding to $D$ is given by
\begin{equation}
  \label{eq:r31}
v_d \equiv c [(3/4\pi)^{1/2} D].
\end{equation}
Assuming $ l_0 \ (\equiv a_0 \chi_0) = 40 (h^{\rm I})^{-1}$ Mpc 
mainly, $D, Q$ 
and $v_d$ were derived for various model parameters. Their values in 
models with a single shell
and double shells are shown in Tables \ref{table1} and \ref{table2}, 
respectively. The values in models with an intermediate self-similar 
region are shown in Table \ref{table3}. In Table \ref{table1} the
values in the case with a moving shell are shown.
  
It is found from Table \ref{table1} that (1) $v_d/v_p$ is smaller than 
0.17 in all cases, \ (2) $v_d$ is approximately proportional to 
$l_0$ (as well as
$v_p$), \ (3) $v_d$ is smaller for smaller $(\Omega_0^{\rm II} -
\Omega_0^{\rm I})$ and $(h^{\rm I} - h^{\rm II})$, \ (4) the
positive cosmological constant plays a role of increasing $v_d$, and \
(5) the influence of the shell motion on $D, Q$ and $v_d$ is
negligibly small. 

If we compare two lines in Table \ref{table2} with
the corresponding ones (the first line and 8th line) in Table
\ref{table1}, the results in the models with a single shell and double
shells are found to be quite consistent. Moreover, if
we compare four lines in Table \ref{table3} with
the corresponding ones (the second line, 4th and 6th line) in Table
\ref{table1}, the results in the models with a single shell and a
self-similar region are found to be similarly consistent. Accordingly
$v_d/v_p$ is $\approx 0.1$ in all models we treated here.

\bigskip

\section{Naive derivation of redshift factors}
\label{sec:level4}

For the dipole anisotropy the maximum difference of temperatures and
redshifts can be seen in a direction ($\phi = \phi_1$) and the inverse 
direction ($\phi = \pi - \phi_1$). Here let us compare the redshifts
in the directions $\phi = 0$ and $\pi$ appearing in the model with a
single shell. The spatial curvature is neglected for simplicity. 
In the $X$ axis we consider six points O, A, B, C, D, and E, for which $X =
\chi_0, \ \chi_1, \ \chi_1 + 2 \chi_0, \ 0, \ -\chi_1,$ and $-\chi_1 
+ 2\chi_0$, as shown in Fig.  \ref{fig4}.

Points B and D have the equal distance from the observer's point O.

The redshifts $z_{\rm rec}(0)$ and $z_{\rm rec}(\pi)$ of the rays from another 
points P and P' at the recombination epoch to the observer at O in the 
directions $\phi = 0$ and $\pi$ are divided into three steps

(P $\rightarrow$ B, B $\rightarrow$ A, A $\rightarrow$ O) \ and \
(P' $\rightarrow$ D, D $\rightarrow$ E, E $\rightarrow$ O),

\noindent respectively. That is, 
\begin{equation}
  \label{eq:n1}
1 + z_{\rm rec}(0) = (1 +z_{\rm PB})(1 +
z_{\rm BA}) (1 + z_{\rm AO}), \ \ 1 + z_{\rm rec}(\pi) = (1 +z_{\rm 
P'D})(1 +z_{\rm DE}) (1 + z_{\rm EO}).
\end{equation}
 Among these three steps, the
first and third steps have equal redshifts evidently : $z_{\rm PB} = 
z_{\rm P'D}, \ z_{\rm AO} = z_{\rm EO}$. In the processes (B
$\rightarrow$ A) and (D $\rightarrow$ E), we have the redshifts due to 
the cosmic expansion, the Doppler shift and the gravitational
redshift. The expansion redshift factors in V$^{\rm II}$ and V$^{\rm
I}$ are 
\begin{equation}
  \label{eq:n2}
[1 + H_0^{\rm II} (2 a_0 \chi_0/c)], \quad [1 + H_0^{\rm I} 
(2 a_0 \chi_0/c)],
\end{equation}
 respectively.  The Doppler shifts at A and D due to 
the relative velocity between V$^{\rm I}$ and V$^{\rm II}$ frames are
\begin{equation}
  \label{eq:n3}
[1 - {1 \over c} (H_0^{\rm I} - H_0^{\rm II})a_0(\chi_1 - \chi_0)], 
\quad [1 - {1 \over c} (H_0^{\rm I} - H_0^{\rm II})a_0(\chi_1 + \chi_0)],
\end{equation}
 respectively. The gravitational redshifts are represented by potentials
$\psi_{\rm BA}$ and $\psi_{\rm DE}$ given by
\begin{equation}
  \label{eq:n4}
\psi_{\rm BA} \equiv {4\pi G (a_0)^3\over c^2}\Big[{1 \over 3}{\rho_0^{\rm I} 
(\chi_1)^3 \over
\chi_1+\chi_0} + {\rho_0^{\rm II} (\chi_1+\chi_0)^2 \chi_0 \over 
\chi_1+\chi_0}\Big],
\end{equation}
\begin{equation}
  \label{eq:n5}
\psi_{\rm DE} \equiv {4\pi G (a_0)^3\over c^2}\Big[{1 \over 3}{\rho_0^{\rm I} 
(\chi_1)^3 \over \chi_1+\chi_0} - {\rho_0^{\rm II} (\chi_1-\chi_0)^2 
\chi_0 \over \chi_1-\chi_0}\Big].
\end{equation}
Accordingly the total redshift factors are
\begin{eqnarray}
  \label{eq:n6}
1 + z_{\rm BA} &=& [1 + H_0^{\rm II} (2 a_0 \chi_0/c)][1 - {1 \over c}
(H_0^{\rm I} - H_0^{\rm II})a_0(\chi_1 - \chi_0)] [1 + \psi_{\rm BA}]\cr
&=& 1 - {a_0 \over c}(H_0^{\rm I} - H_0^{\rm II})\chi_1 + {a_0 \over c}
(H_0^{\rm I} + H_0^{\rm II})\chi_0 + \psi_{\rm BA} \cr
&&- \Big({a_0 \over
c}\Big)^2 H_0^{\rm II} (H_0^{\rm I} - H_0^{\rm II})\chi_0 (\chi_1-\chi_0) 
+ \cdots,
\end{eqnarray}
\begin{eqnarray}
  \label{eq:n7}
1 + z_{\rm DE} &=& [1 + H_0^{\rm I} (2 a_0 \chi_0/c)][1 - {1 \over c}
(H_0^{\rm I} - H_0^{\rm II})a_0(\chi_1 + \chi_0)] [1 + \psi_{\rm DE}] \cr
&=& 1 - {a_0 \over c}(H_0^{\rm I} - H_0^{\rm II})\chi_1 + {a_0 \over c}
(H_0^{\rm I} + H_0^{\rm II})\chi_0 + \psi_{\rm DE} \cr
&&- \Big({a_0 \over
c}\Big)^2 H_0^{\rm I} (H_0^{\rm I} - H_0^{\rm II})\chi_0 (\chi_1+\chi_0) 
+ \cdots.
\end{eqnarray}
The difference of these factors reduces to
\begin{eqnarray}
  \label{eq:n8}
z_{\rm rec}(0) - z_{\rm rec}(\pi) &\simeq& z_{\rm BA} - 
 z_{\rm DE} \cr
 &\simeq&  \psi_{\rm BA} - \psi_{\rm DE} +
\Big({a_0 \over c}\Big)^2 (H_0^{\rm I} - H_0^{\rm II})\chi_0
\Big[(H_0^{\rm I} + H_0^{\rm II})\chi_0 + (H_0^{\rm I} 
- H_0^{\rm II})\chi_1\Big].
\end{eqnarray}
That is, the main terms in $(1 + z_{\rm BA})$ and $(1 + z_{\rm DE})$
cancel.
The ratio ${\cal R}$ of $[z_{\rm rec}(0) - z_{\rm rec}(\pi)]$ 
to the relative velocity between V$^{\rm I}$ and V$^{\rm II}$ is
\begin{eqnarray}
  \label{eq:n9}
{\cal R} \equiv  {z_{\rm rec}(0) - z_{\rm rec}(\pi) \over 
(H_0^{\rm I} - H_0^{\rm
II}) a_0 \chi_0/c} &\simeq& {3 \over 4} \Big[\Omega_0^{\rm II} 
(H_0^{\rm II})^2 (\chi_1+\chi_0) -
\Omega_0^{\rm I} (H_0^{\rm I})^2 (\chi_1-\chi_0)\Big]/(H_0^{\rm I} -
H_0^{\rm II}) \cr
&& + 2 (H_0^{\rm I} + H_0^{\rm II})\chi_0/c + 2 (H_0^{\rm I}
 - H_0^{\rm II})\chi_1/c,
\end{eqnarray}
where we used Eq.(\ref{eq:m9}) for $\rho_0^j$.
This ratio is the counterpart of the ratio of the velocity $v_d$
(corresponding to the dipole moment) to the relative (peculiar)
velocity $v_p$, and these two ratios have comparable values.  

\section{Consistency with the observed large-scale bulk flows}
\label{sec:level5}

As was shown in B of \S 2, the bulk velocities in all positions within 
the region I are equal in the present models, so that the relative 
velocity ($ v_{\rm LG}$) of LG to the cluster frame is only the 
peculiar velocity of LG caused by GA and the nearby superclusters. 
Moreover, as was shown in C of \S 3 the dipole velocity $v_d$
corresponding to the bulk velocity $v_p$ is $\sim 0.1 v_p$, and so the 
total dipole velocity $v_{td}$ of LG to CMB is $v_{td} = v_{\rm LG} +
v_d \ (\sim 0.1 v_p) \cong v_{\rm LG}.$

These conclusions are consistent as follows with \ (1) \ the observed 
velocities
($v_{\rm LG}$) of LG with respect to the cluster frames by \citet{giov98}, 
\citet{dale99} and \citet{riess97}, which are nearly equal to the
total dipole velocity ($v_{td}$) with respect to CMB, and \ (2) \ the
bulk velocities ($v_p$) in the SMAC observation (\citet{hud99}) and
the LP10k observation (\citet{will99}) are $\sim 700$ km sec$^{-1}$ in 
the nearly same directions:

According to \citet{dale99}, we have $v_{\rm LG} = 565 \pm 113$ km
sec$^{-1}$ and $(l, b) = (267^\circ, 26^\circ) \pm 10^\circ$. On the
other hand, $v_{td} = 627 \pm 22$ km sec$^{-1}$ and $(l, b) = 
(276^\circ, 30^\circ)$ \ (\citet{kog93}). Since both directions are
nearly equal, the velocity difference is about 60 km sec$^{-1}$.
This value is comparable with $v_d \ (\sim 0.1 v_p)$, in the
consistent manner with the result in the present models. 

However, they are inconsistent with the observations of \citet{lp94}
and  \citet{coll95} in which $v_{td}$ and $v_{\rm LG}$ are in quite
different directions. At present these observations seem to have been
ruled out (cf. Proceedings of the International Conference on Cosmic 
Flows 1999).     

\section{Concluding remarks}
\label{sec:level6}

In this paper we considered the behaviors of galaxies and light rays
in spherically symmetric inhomogeneous
models consisting of inner and outer homogeneous regions V$^{\rm I}$ 
and V$^{\rm II}$ with $\Omega_0^{\rm I}$ and $\Omega_0^{\rm II} \ 
(> \Omega_0^{\rm I})$ and $H_0^{\rm I}$ and $H_0^{\rm II} \ (< 
H_0^{\rm I})$, respectively, connected by a single shell. In Appendixes 
we treated also the models with double shells and an intermediate 
self-similar
region. It was shown as the result that when we observe the motion of
galaxies at the point O deviated from the center C in V$^{\rm I}$, 
a constant peculiar velocity component $v_p$ appears
in the direction from the center to the observer (C $\rightarrow$ O).
Moreover it was shown that the velocity $v_d$ corresponding to the
dipole anisotropy of  CMB radiation is by a factor $\approx 10$ 
small compared with $v_p$. This result may fit for the observed situation of 
the cosmic flow of cluster galaxies, when the scale of the inner
region and the distance CO are about $200 (h^{\rm I})^{-1}$ and 
$40 (h^{\rm I})^{-1}$ Mpc, respectively, and when $(\Omega_0^{\rm II} 
- \Omega_0^{\rm I})$ is $\ \approx 0.5$ and $(H_0^{\rm I} - H_0^{\rm
II})$ is $\ \approx 18 h^{\rm I}$ km/sec/Mpc \ (or $h^{\rm II}/h^{\rm I} \
\approx 0.82$), respectively.
This difference of the Hubble parameters may be consistent with their
recent values due to nearby and remote observations, because short and 
long distance scales give the Hubble constant of $H_0 \geq 70$ and 
$H_0 \sim 55$, respectively \ \citep{bran98,free97,san97,blan97}.
 A model with multi-shells in the region of $100 \sim
300$ Mpc in which the parameters change stepwise will be better to
reproduce the observed distribution of the Hubble constant. 

It was shown that the present models are consistent with the current
observations of large-scale flows (\citet{giov98}, \citet{dale99},
\citet{riess97}, \citet{hud99} and \citet{will99}), but inconsistent
with other observations (\citet{lp94} and \citet{coll95}) which may
have been ruled out.

It is interesting and important to study the influences of the above
inhomogeneity on the cosmological observations such as the
magnitude-redshift relation of SN1a, the number count of galaxies, 
the time delay for lensed QSOs, and so on. They will be 
quantitatively analyzed and shown in near future.

In this paper the motion of LG due to GA and superclusters in similar 
distances was not treated, while spherical matter distributions on
such scales were analyzed by \citet{hump97}. But the approximation of 
spherical symmetry may not be good to their small-scale matter 
distribution.

\appendix
\section{Models with double shells}

The spacetime is divided into three homogeneous regions V$^{\rm I}$, 
V$^{\rm II}$, and V$^{\rm III}$, as shown in Fig. \ref{fig5}.
 The cosmological parameters in
V$^j$ are shown as $\Omega_0^j, \lambda_0^j$ and so on, where $j =$ 
I, II, III.

Eqs. (\ref{eq:m1}) - (\ref{eq:m3}) and (\ref{eq:m9}) - (\ref{eq:m15})
 hold also in region III as well as in regions I and II.
The junction condition at the second boundary between V$^{\rm II}$ and
V$^{\rm III}$ have similar expressions to those at the first boundary:
\begin{equation}
  \label{eq:a1}
R \equiv a^{\rm II} f^{\rm II} = a^{\rm III} f^{\rm III},
\end{equation}
\begin{equation}
  \label{eq:a2}
\gamma^{\rm III} d(4\pi R^2\sigma)/dt^{\rm III} = \Big[4\pi R^2 
\gamma^2 v\rho \Big]^{\rm III,II}, 
\end{equation}
\begin{equation}
  \label{eq:a3}
d(\gamma^{\rm III} v^{\rm III})/dt^{\rm III} = -\gamma^{\rm III} v^{\rm
III} H^{\rm III} + 2\pi G\sigma - [\gamma^2 v^2 \rho]^{\rm III}/\sigma,
\end{equation}
\begin{equation}
  \label{eq:a4}
dt^{\rm II}/dt^{\rm III} = \gamma^{\rm II}/\gamma^{\rm III},
\end{equation}
\begin{equation}
  \label{eq:a5}
\gamma^{\rm II}(f'^{\rm II} v^{\rm II} + H^{\rm II}R) = \gamma^{\rm III}
(f'^{\rm III} v^{\rm III} + H^{\rm III} R),
\end{equation}
and
\begin{equation}
  \label{eq:a6}
[\gamma(f' +vHR)]^{\rm III,II} = - 4\pi G\sigma R,
\end{equation}
where $[\Phi]^{\rm III,II} \equiv \Phi^{\rm III} - \Phi^{\rm II}$.

Similarly to Eqs. (\ref{eq:m19}) and (\ref{eq:m20}), moreover, we have
\begin{equation}
  \label{eq:a7}
{\rho_0^{\rm III} \over \rho_0^{\rm II}} = \Big({y_i^{\rm III} 
\over y_i^{\rm II}}\Big)^3 (1 + \epsilon y_i^{\rm II}),
\end{equation}
\begin{equation}
  \label{eq:a8}
\Big({H_0^{\rm III} \over H_0^{\rm II}}\Big)^2 = \Big({y_i^{\rm III} 
\over y_i^{\rm II}}\Big)^3 + \lambda_0^{\rm II} \Big[1 - \Big({y_i^{\rm III} 
\over y_i^{\rm II}}\Big)^2 \Big] + {y_i^{\rm III} 
\over y_i^{\rm II}}\Big)^3  \Big[1 - (1 + \epsilon) {y_i^{\rm II} 
\over y_i^{\rm III}}\Big] \Omega_0^{\rm II},
\end{equation}
where $\epsilon \approx 1$.

As for light rays, the equations in V$^{\rm I}$ and V$^{\rm II}$ are
same as those in Sec. 3, and equations in V$^{\rm III}$ are common
with those in V$^{\rm II}$. At the second boundary ($\eta = \eta_2, 
\ \chi = \chi_2$) we have    
\begin{equation}
  \label{eq:a9}
\Big({a_0 \over a_2}\Big)^{\rm II} w_2^{\rm II} = \Big({a_0 
\over a_1}\Big)^{\rm III} w_2^{\rm III}, 
\end{equation}
\begin{equation}
  \label{eq:a10}
d^{\rm II} = d^{\rm III}, 
\end{equation}
where $a_2 = a(\eta_2)$.

Under the assumption that the shells are comoving (i.e., $v^{\rm I} 
= v^{\rm II} = v^{\rm III} = 0$), we obtain the following redshift 
formulas:
\begin{equation}
  \label{eq:a11}
1 + z_1^{\rm I} = {1 \over y^{\rm I}} (\eta_1^{\rm I})
\end{equation}
and in V$^{\rm II}$ we have 
\begin{equation}
  \label{eq:a12}
{1 + z_2^{\rm II} \over 1 + z_1^{\rm II}} = {y^{\rm II} 
(\eta_1^{\rm II}) \over y^{\rm II} (\eta_2^{\rm II})}.
\end{equation}
\begin{equation}
  \label{eq:a13}
{1 + z_{\rm rec}^{\rm III} \over 1 + z_2^{\rm III}} = {y^{\rm III} 
(\eta_2^{\rm III}) \over y^{\rm III} (\eta_{\rm rec}^{\rm III})}.
\end{equation}
From Eqs. (\ref{eq:m6}) and (\ref{eq:a4}) we have two equations 
${z}_1^{\rm I} = {z}_1^{\rm II}$ and ${z}_2^{\rm II} = 
{z}_2^{\rm III}$. Moreover, $(\eta_1^{\rm II}, \chi_1^{\rm II})$ 
are related to $(\eta_1^{\rm I}, \chi_1^{\rm I})$  using Eqs. 
(\ref{eq:r22}) and (\ref{eq:r23}), and $(\eta_2^{\rm III}, 
\chi_2^{\rm III})$ are related to $(\eta_2^{\rm II}, \chi_2^{\rm II})$
using the similar equations
\begin{equation}
  \label{eq:a14}
a_0^{\rm II} y^{\rm II} (\eta_2^{\rm II}) \sinh \chi_2^{\rm II} = 
a_0^{\rm III} y^{\rm III} 
(\eta_2^{\rm III}) \sinh \chi_2^{\rm III} 
\end{equation}
and 
\begin{equation}
  \label{eq:a15}
a_0^{\rm II} \int_0^{\eta_2^{\rm II}} y^{\rm II} (\eta^{\rm II}) 
d\eta^{\rm II} = a_0^{\rm III} \int_0^{\eta_2^{\rm III}} y^{\rm III} 
(\eta^{\rm III}) d\eta^{\rm III}.
\end{equation}

First we consider the virtual observer in the center $(\chi = 0)$. 
Then we have
\begin{equation}
  \label{eq:a16} 
1 + \bar{z}_1^{\rm I} = {1 \over y^{\rm I}(\bar{\eta}_1^{\rm I})}
\end{equation}
\begin{equation}
  \label{eq:a17}  
{1 + \bar{z}_2^{\rm II} \over 1 + \bar{z}_1^{\rm II}} = {y^{\rm II}
(\bar{\eta}_1)^{\rm II} \over y^{\rm II}(\bar{\eta}_2)^{\rm II}},
\end{equation}
\begin{equation}
  \label{eq:a18}  
{1 + \bar{z}_{\rm rec}^{\rm III} \over 1 + \bar{z}_2^{\rm III}} 
= {y^{\rm III}(\bar{\eta}_2)^{\rm III} \over y^{\rm III}(\bar{\eta}_{\rm 
rec})^{\rm III}},
\end{equation}
where $\bar{\eta}_0^{\rm I} - \bar{\eta}_1^{\rm I} = \chi_1^{\rm I} $ and 
$\bar{\eta}_2^{\rm II} - \bar{\eta}_1^{\rm II} = \chi_2^{\rm II} - 
\chi_1^{\rm II}$.
Accordingly, we can determine $\chi_1^{\rm I}, \chi_2^{\rm II}$ and 
$\bar{\eta}_{\rm rec}^{\rm III}$
by specifying his redshifts $\bar{z}_1^{\rm I},\ \bar{z}_2^{\rm II}$ 
and $\bar{z}_{\rm rec}^{\rm III} \ (= 10^3 -1)$, as in Sec. 3.

For an observer at O (with $\eta = \eta_0$ and $\chi = \chi_0$), we
obtain 
\begin{equation}
  \label{eq:a19}   
1 + z_{\rm rec}^{\rm III} = {y^{\rm II} (\eta_1^{\rm II}) y^{\rm 
III} (\eta_2^{\rm III})\over y^{\rm I} (\eta_1^{\rm I}) y^{\rm II} 
(\eta_2^{\rm II}) y^{\rm III} (\eta_{\rm rec}^{\rm III})},
\end{equation}
where $\eta_1^{\rm II}, \eta_2^{\rm II}, \eta_2^{\rm III}$ and 
$\eta_{\rm rec}^{\rm III}$ depend on the angle $\phi$, contrary to
those in Eq. (\ref{eq:a18}). When we fix $\bar{z}_1^{\rm I},\ 
\bar{z}_2^{\rm II}$ and $\bar{z}_{\rm rec}^{\rm III}$ and determine 
$\chi_1^{\rm I}, \chi_2^{\rm II}$ and 
$\bar{\eta}_{\rm rec}^{\rm III}$, the final
redshift $z_{\rm rec}^{\rm III}$ can be obtained as a function of
angle $\phi$ for ${\eta}_{\rm rec}^{\rm III} = \bar{\eta}_{\rm 
rec}^{\rm III}$. It is to be noted that ${\eta}_{\rm rec}^{\rm III}$
is independent of the existence of the two shells.

\section{Models with an intermediate self-similar region}

The line-element is expressed in the form (Tomita 1995, 1996)
\begin{equation}
  \label{eq:b1}
ds^2 = -c^2 dt^2 + S^2(t,r)\Biggl\{ {(1+rS'/S)^2 \over 1-k\alpha(r)r^2} dr^2
+r^2d\Omega^2 \Biggr\},
\end{equation}
where 
\begin{equation}
  \label{eq:b2}
\alpha (r)/\alpha_0 =1/(r_1)^2,\ 1/r^2, \ {\rm and} \ 1/(r_2)^2
\end{equation}
for the inner homogeneous region V$^{\rm I}$, the self-similar region 
V$^{\rm II}$, and the outer homogeneous region V$^{\rm III}$,
respectively, which are shown in Fig. \ref{fig7}. Assuming the 
matter pressure-free and comoving, the
solutions are described using the Tolman solution. The two boundaries
$r = r_1$ and $r_2$ are exactly comoving. The scale factor $S$ 
in the case $\Lambda = 0$ is given by
\begin{equation}
\label{eq:b3}
S/S_0 = {\Omega_0 \over 2(1-\Omega_0)} (\cosh \eta -1),
\end{equation}
and
\begin{equation}
\label{eq:b4}
\Omega_0 = (cH_0^{-1}/S_0)^3, \quad H_0 \equiv [(\partial 
S/\partial t)/S]_0.
\end{equation}

The time variable $t$ is in the inner, self-similar, and
outer regions expressed as 
\begin{equation}
\label{eq:b5}
H_0t = {\Omega_0 \over 2(1-\Omega_0)^{3/2}} (\sinh \eta -\eta),
\end{equation}
\begin{equation}
\label{eq:b6}
\xi \equiv {ct \over r} = {c{H_0}^{-1} (r_1)^{-1} \Omega_0 \over
2(1-\Omega_0)^{3/2}}(\sinh \eta -\eta),
\end{equation}
\begin{equation}
\label{eq:b7}
H_0 t = {r_2 \over r_1}{\Omega_0 \over [2(1- \Omega_0)^{3/2}]} (\sinh
\eta - \eta).
\end{equation}
respectively.he constant $\alpha_0$ and $r_1$ are given by
\begin{equation}
\label{eq:b8}
\sqrt{\alpha_0} = {4 \sqrt{1-\Omega_0}\over
{\Omega_0}^2(1+\bar{z}_1)}\Biggl[1-\Omega_0+{\Omega_0 \over 2}(1+\bar{z}_1)+
\Biggl({\Omega_0
\over 2}-1\Biggr)\sqrt{1+\Omega_0\bar{z}_1}\Biggr].
\end{equation}
and 
\begin{equation}
\label{eq:b9}
r_1 =[(\Omega_0)^{1/3}/(1-\Omega_0)^{1/2}]\sqrt{\alpha_0}.
\end{equation}
If we define the local Hubble parameter $H$ (in the transverse
direction) in the $t = t_0$ hypersurface as
\begin{equation}
\label{eq:b10}
H \equiv (\dot S/S)_{t=t_0} = [(1/t)\xi S_{,\xi}/S]_{t=t_0}, 
\end{equation}
$H$ in V$^{\rm I}$ is equal to $H_0$ and we have in V$^{\rm II}$ and 
V$^{\rm I}$
\begin{equation}
\label{eq:b11}
Ht_0 = {\sinh \eta (\sinh \bar\eta - \bar\eta) \over (\cosh \bar\eta
-1)^2},
\end{equation}
where ${\bar \eta} \equiv \eta_{t = t_0}$  satisfies
\begin{equation}
\label{eq:b12}
H_0 t_0 = \Big({r \over r_1}, {r_2 \over r_1}\Big){\Omega_0 
\over [2(1- \Omega_0)^{3/2}]} (\sinh \bar \eta - \bar \eta).
\end{equation}
for (V$^{\rm II}$, V$^{\rm III}$, respectively.
Present matter densities in the three regions are
\begin{equation}
\label{eq:b13}
\rho_{0j} = {3 {H_0}^2 \Omega_0 \over 8\pi G}\Big[1, {1 \over 3}
\biggl({r_1 \over r}\biggr)^2 \biggl(1-\xi S_{,\xi}/S\biggr)^{-1},
\biggl({r_1 \over r_2}\biggr)^2\Big] 
\end{equation}
for $j = [$I, II, III$]$, respectively, and $\Omega_{0j}$ is defined 
as
\begin{equation}
\label{eq:b14}
\Omega_{0j} \equiv \rho_{0j}(t_0)/\Biggl({3H^2 \over 8\pi G}\Biggr).
\end{equation}

The behavior of light rays and redshift formulas were derived in the
previous paper\citep{tom96}. If we specify the redshifts 
$\bar{z}_1,\ \bar{z}_2$ and $\bar{z}_{\rm rec} (= 10^3 -1)$, the radii 
$r_1$ and $r_2$ of two boundaries and the value of $\eta_{\rm rec} (= 
\bar{\eta}_{\rm rec})$ are determined, and we can calculate $z_{\rm
rec}$ as a function of $\phi$, which is the angle between the incident 
direction and the $X$ axis.

In the following we present additional explanations and corrections
for the formulas in the Appendix A of the previous paper\citep{tom96} :
In V$^{\rm I}$ and V$^{\rm III}$ we have
\begin{equation}
\label{eq:b15}
k^r = \pm (S_0/S^2)\{[1+\alpha_0(r/r_1)^2][(w_1)^2-d^2/(S_0r)^2]\}^{1/2},
\end{equation}
\begin{equation}
\label{eq:b16}
k^r = - (S_0/S^2)\{[1+\alpha_0(r/r_2)^2][(w_2)^2-d^2/(S_0r)^2]\}^{1/2},
\end{equation}
respectively. The second ray in  V$^{\rm I}$ has 
\begin{equation}
\label{eq:b17}
h_1 =\cosh^{-1}\Biggl\{\Biggl[1+(\sqrt{\alpha_0}r_0/r_1)^2\Biggr]^{1/2}
/h_0\Biggr\} +\{(\eta-\eta_0), -(\eta-2\eta_m+\eta_0)\}
\end{equation}
for $\{\eta \geq \eta_m, \eta \leq \eta_m\}$, respectively.
In this paper the incoming angle is 
\begin{equation}
\label{eq:b18}
\phi = \sin^{-1} \Biggl({d/w_1 \over S_0r_0}\Biggr) (\equiv \phi_1)
\quad {\rm and} \quad \pi - \phi_1.
\end{equation}
Auxiliary functions V$^{\rm III}$ are expressed as 
\begin{equation}
\label{eq:b19}
M(\xi) \equiv {1 \over N(\xi)}\Biggl\{(M_0)^2-2d^2\int^\xi_{\xi_1}d\xi
{N(\xi) \over S^3}\Biggl[SS_{,\xi} + \xi\Bigl(1+\alpha_0+SS_{,\xi\xi}-
(S_{,\xi})^2\Bigr)\Biggr]\Biggr\}^{1/2},
\end{equation}
\begin{equation}
\label{eq:b20}
\zeta(\xi) \equiv N(\zeta)/[\sqrt{1+\alpha_0}\xi],
\end{equation}
\begin{equation}
\label{eq:b21}
{M \over d} ={2(1-\Omega_0) \over \Omega_0 S_0}{\cosh \eta -1 \over 2(1-\cosh
\eta)+\eta \sinh \eta}\Bigl({M_0 \over d} -2I\Bigr),
\end{equation}
and
\begin{equation}
\label{eq:b22}
\zeta = {\sqrt{\alpha_0} \over \sqrt{1+\alpha_0}}{2(1-\cosh \eta)+\eta \sinh 
\eta \over (\cosh \eta -1)(\sinh \eta - \eta)}.
\end{equation}

The constants $w_1$ and $w_2$ are connected using the junction
conditions $(k^0)^{\rm I} = (k^0)^{\rm II}$ at $r = r_1$ and 
$(k^0)^{\rm II} = (k^0)^{\rm III}$ at $r = r_2$.

The author thanks N. Sugiyama and N. Sakai for helpful discussions,
and a referee for kind comments about past related
works. This work was supported by Grant-in Aid for Scientific Research 
(No. 10640266) from the Ministry of Education, Science, Sports and
Culture, Japan.


\clearpage

\figcaption[bulk1.eps]{A model with a single shell. $z$ and $\bar{z}$ are the
 redshifts for observers at O and C.  \label{fig1}}

\figcaption[bulk2.eps]{A schematic diagram of Hubble and density parameters.
 A solid and dotted lines denote $H_0$ and $\Omega_0$, respectively.
  \label{fig2}}

\figcaption[bulk3.eps]{Components of the bulk velocity.  \label{fig3}}

\figcaption[bulk4.eps]{A diagram for the naive derivation.
  \label{fig4}}

\figcaption[bulk5.eps]{A model with double shells. \label{fig5}}

\figcaption[bulk6.eps]{A schematic diagram of Hubble and density 
parameters in a
 model with double shells. A solid and dotted lines denote
 $H_0$ and $\Omega_0$, respectively.  \label{fig6}}

\figcaption[bulk7.eps]{A model with the self-similar region. \label{fig7}} 

\clearpage
\figcaption[bulk8.eps]{A schematic diagram of Hubble and density 
parameters in a
 model with the self-similar region. A solid and dotted lines denote
 $H_0$ and $\Omega_0$, respectively.  \label{fig8}}

\clearpage

\begin{table}
\caption{Dipole and quadrupole moments and the velocity $v_d$ in the
single-shell models.
\label{table1}}
\begin{tabular}{ccccccccc}
$\Omega_0^{\rm I}$&$\Omega_0^{\rm II}$&$\lambda_0^{\rm I}$&
$\lambda_0^{\rm II}$&$h^{\rm I}$&$h^{\rm II}/h^{\rm I}$&$D \ (\times
10^4)$&$Q \ (\times 10^5)$&$v_d$ \ (km/sec) \\
\tableline
0.2& 0.56 & 0.0 & 0.0 &0.7& 0.82 & 5.56 &-2.87& 81.8\tablenotemark{1} \\
0.2& 0.88 & 0.0 & 0.0 &0.7& 0.82 & 8.27 &-4.27&121.5\tablenotemark{1} \\
0.2& 0.56 & 0.0 & 0.0 &0.7& 0.90 & 5.38 &-2.79& 79.1\tablenotemark{1} \\
0.2& 0.88 & 0.0 & 0.0 &0.7& 0.90 & 8.21 &-4.23&120.7\tablenotemark{1} \\
0.3& 0.56 & 0.0 & 0.0 &0.7& 0.82 & 3.90 &-2.04& 57.3\tablenotemark{1} \\
0.3& 0.88 & 0.0 & 0.0 &0.7& 0.82 & 6.66 &-3.39& 97.9\tablenotemark{1} \\
0.2& 0.56 & 0.672& 0.430 &0.7& 0.82 & 7.92 &-3.74&111.4\tablenotemark{1} \\
0.2& 0.56 & 0.0 & 0.0 &0.7& 0.82 & 5.55 &-1.91& 81.6\tablenotemark{2} \\
0.2& 0.88 & 0.0 & 0.0 &0.7& 0.82 & 8.21 &-2.68&120.7\tablenotemark{2} \\
0.2& 0.56 & 0.0 & 0.0 &0.7& 0.82 & 8.07 &-6.35&118.6\tablenotemark{3} \\
0.2& 0.88 & 0.0 & 0.0 &0.7& 0.82 &12.31 &-9.62&181.0\tablenotemark{3} \\
0.2& 0.56 & 0.0 & 0.0 &0.7& 0.82 & 5.57 &-2.87& 81.8\tablenotemark{4} \\
0.2& 0.88 & 0.0 & 0.0 &0.7& 0.82 & 8.27 &-4.27&121.6\tablenotemark{4} \\
\end{tabular}
\tablenotetext{1}{$\bar{z}_1^{\rm I} = 0.067, \ l_0 = 40 (h^{\rm I})^{-1}.$}
\tablenotetext{2}{$\bar{z}_1^{\rm I} = 0.1, \ l_0 = 40 (h^{\rm I})^{-1}.$}
\tablenotetext{3}{$\bar{z}_1^{\rm I} = 0.067, \ l_0 = 60 (h^{\rm I})^{-1}.$}
\tablenotetext{4}{$\bar{z}_1^{\rm I} = 0.067, \ l_0 = 40 (h^{\rm I})^{-1},$ \ 
and the shell velocity is 200 km/sec. }
\end{table}

\clearpage
\begin{table}
\caption{Dipole and quadrupole moments and the velocity $v_d$ in the
double-shell models in the case of $\lambda_0^j = 0$ and $l_0 = 40
(h^{\rm I})^{-1}$.
\label{table2}}
\begin{tabular}{ccccccccc}
$\Omega_0^{\rm I}$&$\Omega_0^{\rm II}$&$\Omega_0^{\rm III}$&
$h^{\rm I}$&$h^{\rm II}/h^{\rm I}$&$h^{\rm III}/h^{\rm I}$&$D \ (\times
10^4)$&$Q \ (\times 10^5)$&$v_d$ \ (km/sec) \\
\tableline
0.2& 0.36 & 0.56 &0.7&0.92& 0.82 & 5.55 &-2.37& 81.6\tablenotemark{1}\\
0.2& 0.36 & 0.56 &0.7&0.92& 0.82 & 5.52 &-1.44& 81.1\tablenotemark{2}\\
\end{tabular}
\tablenotetext{1}{$\bar{z}_1^{\rm I} = 0.067, \ \bar{z}_1^{\rm II} 
= 0.1.$}
\tablenotetext{2}{$\bar{z}_1^{\rm I} = 0.1, \ \bar{z}_1^{\rm II} =
0.2.$}
\end{table}

\clearpage
\begin{table}
\caption{Dipole and quadrupole moments and the velocity $v_d$ in the
model with a self-similar region in the case of $\lambda_0^j = 0$ and
$l_0 = 40 (h^{\rm I})^{-1}$.
\label{table3}}
\begin{tabular}{ccccccc}
$\Omega_0^{\rm I}$&$\Omega_0^{\rm III}$&
$h^{\rm I}$&$h^{\rm III}/h^{\rm I}$&$D \ (\times
10^4)$&$Q \ (\times 10^5)$&$v_d$ \ (km/sec) \\
\tableline
0.2& 0.89 &0.7&0.81& 8.40 &-4.41& 123.4\tablenotemark{1}\\
0.2& 0.72 &0.7&0.83& 7.40 &-2.61& 108.8\tablenotemark{2}\\
0.3& 0.91 &0.7&0.84& 8.08 &-4.11& 118.8\tablenotemark{1}\\
0.3& 0.80 &0.7&0.86& 6.94 &-2.25& 102.0\tablenotemark{2}\\
\end{tabular}
\tablenotetext{1}{$\bar{z}_1^{\rm I} = 0.067, \ \bar{z}_1^{\rm II} 
= 0.1.$}
\tablenotetext{2}{$\bar{z}_1^{\rm I} = 0.1, \ \bar{z}_1^{\rm II} =
0.2.$}
\end{table}

\end{document}